\documentclass[11pt,twoside]{article}

\usepackage{asp2006}
\usepackage{epsf}
\usepackage{graphicx}
\usepackage{lscape}

\begin{document}
\title{Signatures of Granulation in the Spectra of K-Dwarfs}   
\author{I. Ram\'irez,\altaffilmark{1} C. Allende Prieto,\altaffilmark{1} D. L. Lambert,\altaffilmark{1} \& M. Asplund\altaffilmark{2}}   
\affil{\altaffilmark{1} Department of Astronomy, University of Texas at Austin \\ \altaffilmark{2}  Max Planck Institute for Astrophysics}    

\begin{abstract} 
Very high resolution ($R>150,000$) spectra of a small sample of nearby K-dwarfs have been acquired to measure the line asymmetries and central wavelength shifts caused by convective motions present in stellar photospheres. This phenomenon of granulation is modeled by 3D hydrodynamical simulations but they need to be confronted with accurate observations to test their realism before they are used in stellar abundance studies. We find that the line profiles computed with a 3D model agree reasonably well with the observations. The line bisectors and central wavelength shifts on K-dwarf spectra have a maximum amplitude of only about 200\,m\,s$^{-1}$ and we have been able to resolve these granulation effects with a very careful observing strategy. By computing a number of iron lines with 1D and 3D models (assuming local thermodynamic equilibrium), we find that the impact of 3D-LTE effects on classical iron abundance determinations is negligible.
\end{abstract}



\section{Context}

K-dwarfs are ideal for Galactic chemical evolution studies because they are not biased by stellar death (lifetimes of K-dwarfs are greater than the age of the Galaxy). These cool dwarf stars have convective envelopes and, therefore, they experience granulation (small-scale convective motions associated with temperature and density fluctuations) in their photospheres. However, classical model atmospheres of K-dwarfs (e.g., ATLAS, MARCS) are static and homogeneous, i.e., incompatible with granulation, yet they are extensively used in stellar abundance work.

Granulation affects line profiles and line strengths, which are the basis for chemical abundance determinations from stellar spectra. Severe inconsistencies in K-dwarf abundance studies for key elements such as Fe and O have been reported recently in the literature (e.g., Morel \& Micela 2004, Schuler et al. 2006, Ram\'irez et al. 2007), which may originate in the inadequacy of standard spectral line-formation calculations that use classical model atmospheres.

We aim at detecting and quantifying the signatures of granulation in very high resolution spectra of K-dwarfs. The Doppler shifts introduced by the convective motions result in asymmetric absorption line-profiles whose central wavelengths are shifted with respect to their rest values (e.g., Allende Prieto et al. 2002, Dravins et al. 1981, Dravins 1987, Gray 1982, 2005). In contrast, classical model atmospheres predict non-shifted and perfectly symmetric lines. Furthermore, we use a state-of-the-art three-dimensional radiative-hydrodynamical model atmosphere to explore the impact of granulation on standard abundance studies of K-dwarfs. In this paper, we present preliminary results from our study.

\section{Observations and modeling}

A small sample of bright K-dwarfs has been observed with the 2dcoud\'e spectrograph (Tull et~al. 1995) on the 2.7-m Telescope at McDonald Observatory. The wavelength coverage of these data is complete in the interval $\lambda\lambda=5580-7800$\,\AA\ and the spectral resolution ($R=\lambda/\Delta\lambda$) ranges from 150,000 to 210,000. Very high signal-to-noise ratios ($S/N>300$) were achieved by carefully coadding several exposures of the same object. Observations from the Hobby-Eberly Telescope are also being used in this study. Details on the data reduction and post-reduction processing will be given in a forthcoming publication (Ram\'irez et~al. 2008a). Similar high quality data for a small sample of stars across the HR diagram will be analyzed in a future study (Ram\'irez et~al. 2008b).

A three-dimensional radiative-hydrodynamical model atmosphere of parameters $T_\mathrm{eff}=4820$\,K, $\log g=4.5$, and [Fe/H]=0 was computed using the prescription described in Stein \& Nordlund (1998, see also Asplund et al. 2000). Several absorption lines were calculated using the 3D model and have been used to determine theoretical line-bisectors and central wavelength shifts (see Figs. 1 and 2). In particular, in Fig. 2 we show the relation between central wavelength shift and equivalent width; we used 10 \ion{Fe}{i} lines of different wavelength and EP values to construct this trend, which has an intrinsic line-to-line scatter of only 10~m\,s$^{-1}$. Details will be given in Ram\'irez et~al. (2008a).

\section{Results}

By looking at a few iron lines in our observed spectra we find that the
theoretical line profiles predicted by the 3D model are reasonably consistent
with the observations. Fig.~1 shows an example of this. Furthermore, the central
wavelength shifts predicted by the 3D model are in remarkable agreement with the
observations, as it is shown in Fig.~2 for the case of HIP~86400, our sample star with parameters closest to that of the 3D model. Although the scatter is large, the standard deviation of that distribution ($\sim$130~m~s$^{-1}$) is fully explained by errors in the dispersion solution ($\sim$25~m~s$^{-1}$), merging of spectral orders and setups ($\sim$50~m~s$^{-1}$), uncertain laboratory wavelengths ($<\sim$75~m~s$^{-1}$), and finite signal-to-noise ratios ($\sim$90~m~s$^{-1}$). 

These results validate our 3D model atmosphere and allow us to use it confidently in stellar abundance determinations.

\begin{figure}
\centering
\includegraphics[width=4.0in]{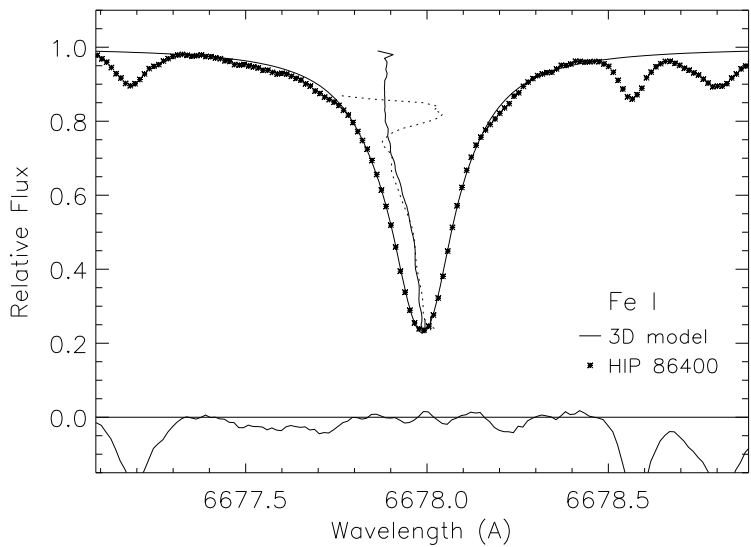}
\caption{\ion{Fe}{i} line profile observed in the spectrum of HIP~86400 (stars). The synthetic 3D line-profile is shown with the solid line. The observed and predicted line-bisectors (midpoints of the horizontal line segments between the wings of the line) are shown within the spectral line and have been expanded for clarity. The maximum amplitude of this bisector is only about 200~m\,s$^{-1}$. Note that the observed bisector (dotted line) becomes severely affected by small blends as one approaches the continuum level. Residuals (observation--model), expanded by a factor of 2, are shown at the bottom.}
\end{figure}

Using the small set of \ion{Fe}{i} and \ion{Fe}{ii} lines computed with the 3D model and comparing their strengths with those predicted by a standard spectrum synthesis procedure using a classical 1D Kurucz model, we find that the impact of 3D effects on the abundance of Fe is very small (see Fig.~3) and it is unlikely that 3D effects will solve the problems reported by the studies cited in Sect.~1.

\begin{figure}
\centering
\includegraphics[width=3.4in]{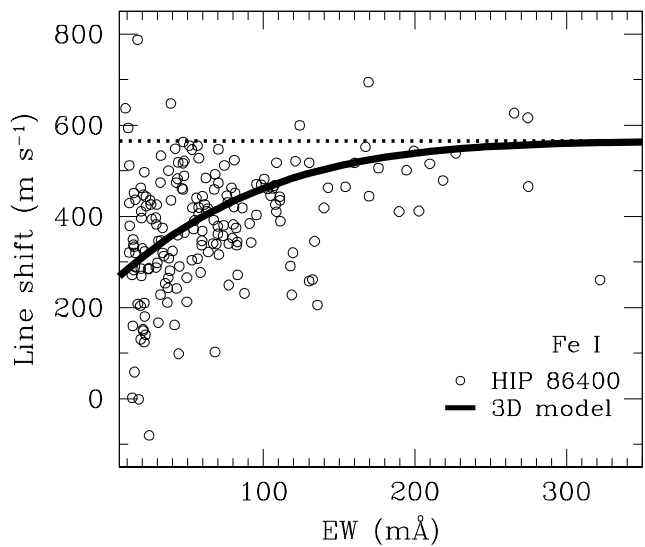}
\caption{Central wavelength shifts for about 180 \ion{Fe}{i} lines with well determined rest wavelengths, as measured in our high resolution spectrum of HIP~86400 (circles) and predicted by the 3D model (solid line). A K-S test shows that the distribution of points around the theoretical line corresponds to the expected random scatter with a 90\% confidence level.}
\end{figure}

\begin{figure}[ht]
\centering
\includegraphics[width=3.2in]{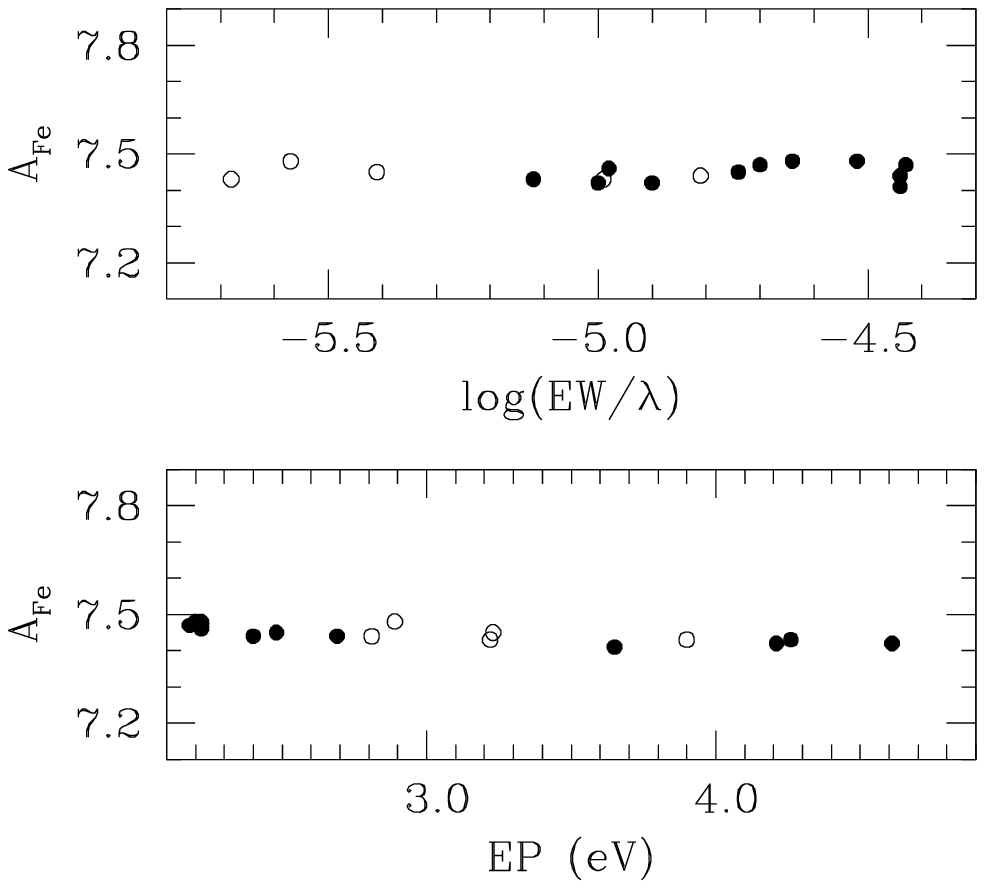}
\caption{Here, the abundance of Fe was determined using a standard technique, i.e., ATLAS 1D model atmosphere (e.g., Kurucz 1979) and the spectrum synthesis code MOOG (Sneden 1973) with the equivalent widths predicted by the 3D model. Abundances from \ion{Fe}{i} (filled circles) and \ion{Fe}{ii} (open circles) lines are shown as a function of reduced equivalent width ($\mathrm{REW}=\log(\mathrm{EW}/\lambda)$) and excitation potential (EP). The equivalent widths predicted by the 3D model were computed with a unique Fe abundance ($A_\mathrm{Fe}=7.45$). When used in the 1D analysis, the resulting abundance was $A_\mathrm{Fe}=7.45\pm0.02$ for \ion{Fe}{i} lines and $A_\mathrm{Fe}=7.44\pm0.02$ for \ion{Fe}{ii} lines. The fact that the abundances from \ion{Fe}{i} and \ion{Fe}{ii} lines are consistent (within 0.01~dex) and no significant trends with EP or REW are introduced in the 1D analysis suggests that the impact of 3D effects on the Fe abundance determination in K-dwarfs is negligible.}
\end{figure}

\acknowledgements 
This work was supported in part by the Robert~A.~Welch Foundation of Houston, Texas. CAP's research is funded by NASA (NAG5-13057 and NAG5-13147).


\end{document}